\documentclass[aps,pra,reprint,superscriptaddress,longbibliography,%
amsmath,amssymb,floatfix]{revtex4-2}

\usepackage{graphicx}
\usepackage{bm}
\usepackage{xcolor}
\usepackage[colorlinks=true,linkcolor=blue!50!black,citecolor=blue!50!black,%
urlcolor=blue!50!black]{hyperref}

\graphicspath{{figures/}}

\newif\ifshowreview \showreviewfalse
\DeclareRobustCommand{\revnote}[1]{\ifshowreview{\,[\textbf{Reviewer:} #1]}\fi}

\definecolor{revisiongreen}{RGB}{0,110,75}
\newif\ifshownewchanges \shownewchangestrue
\DeclareRobustCommand{\newchange}[1]{{#1}}

\newif\ifshowcpsuggestions \showcpsuggestionsfalse
\newcommand{\cpsuggestion}[2]{\ifshowcpsuggestions\par\smallskip\noindent
  {\textbf{Conditional CP suggestion #1 (not manuscript text).}\ #2\par}
  \smallskip\fi}

\newcommand{\Hc}{\mathcal{H}}                  
\newcommand{\Ec}{\mathcal{E}}                  
\newcommand{\Uc}{\mathcal{U}}                  
\newcommand{\Pbar}{\overline{P}}               
\newcommand{\ket}[1]{\left|#1\right\rangle}
\newcommand{\bra}[1]{\left\langle#1\right|}
\newcommand{\T}{\mathrm{T}}
\IfFileExists{dsfont.sty}{\usepackage{dsfont}\newcommand{\id}{\mathds{1}}}{%
  \IfFileExists{bbm.sty}{\usepackage{bbm}\newcommand{\id}{\mathbbm{1}}}{%
    \newcommand{\id}{\mathrm{1\!\!1}}}}

\newcommand{\Pz}{P^0}
\newcommand{\Pzb}{\bar{P}^0}
\newcommand{\Kz}{K^0}
\newcommand{\Kzb}{\bar{K}^0}
\newcommand{\Bd}{B_d^0}
\newcommand{\Bdb}{\bar{B}_d^0}
\newcommand{\Bs}{B_s^0}
\newcommand{\Bsb}{\bar{B}_s^0}

\begin{document}

\title{Quantum simulation of neutral-meson mixing and decay on a reprogrammable photonic processor}

\author{Mohd Abu Bakar Siddique}
\email[Contact author: ]{[abubakar39@iut-dhaka.edu]}
\affiliation{Islamic University of Technology (IUT), Board Bazar, Gazipur 1704, Bangladesh}
\author{Abrar Al Shadid Abir}
\affiliation{Islamic University of Technology (IUT), Board Bazar, Gazipur 1704, Bangladesh}
\author{A. K. M. Rakib}
\affiliation{Islamic University of Technology (IUT), Board Bazar, Gazipur 1704, Bangladesh}
\author{Mohammad Abrar Kabir}

\affiliation{Islamic University of Technology (IUT), Board Bazar, Gazipur 1704, Bangladesh}
\author{Aseer Imad Keats}
\affiliation{Islamic University of Technology (IUT), Board Bazar, Gazipur 1704, Bangladesh}
\author{Rakibul Hasan Sagor}
\affiliation{Islamic University of Technology (IUT), Board Bazar, Gazipur 1704, Bangladesh}

\date{\today}

\begin{abstract}
Neutral mesons oscillate into their antiparticles while decaying, so their dynamics is generated by a non-Hermitian effective Hamiltonian which a quantum simulator restricted to unitary operations cannot implement directly. We propose a reprogrammable photonic processor that simulates this open-system evolution through dilation and analyze it at the circuit level. The effective Hamiltonian is reconstructed exactly from measured lifetimes, mass and width differences, and the parameter $|q/p|$ that quantifies charge-parity (CP) violation in mixing. Through a singular-value decomposition and a single ancilla that encodes the total decay probability, its propagator is embedded in a four-mode unitary transformation, which is compiled at several time steps into a rectangular interferometer on a hexagonal mesh of tunable Mach--Zehnder units. Circuit simulations that include coupler, propagation, and bend losses show that one fixed device, reprogrammed only through its control voltages, reproduces the flavor oscillation and decay of neutral kaons and of $B_d$ and $B_s$ mesons, whose mixing parameter $x=\Delta m/\Gamma$ ranges from 0.77 to 26.9. Over 500 independently programmed time steps per system, the mean fidelity relative to the compiled transformation\revnote{Replace the phrase ``the mean transformation fidelity'' with: ``the mean fidelity relative to the compiled transformation''.} lies between 0.99979 and 0.99982, and no step falls below 0.9996. Because every optical path crosses nine tunable units, a uniform insertion loss reduces only the transmitted power, whereas unit-to-unit loss variations lower the fidelity quadratically, $1-\langle F\rangle\simeq0.08\,\sigma^2$ for a loss standard deviation $\sigma$ in decibels. A leading-order analysis traces this coefficient to the routing of light through the mesh rather than to the simulated dynamics, reproduces the fitted value to within 2\%, and explains why it is the same for all three mesons. A mean fidelity of 0.999 tolerates $\sigma\approx0.11$~dB.The dedicated ancillary outputs encode the total decay probability without postselection on the ancilla outcome and the reported probabilities are normalized to the collected optical power.
\end{abstract}

\maketitle

\section{Introduction}
\label{sec:introduction}

From the inception of quantum simulation era, emulating the dynamics of the subatomic particles has been a prominent subject of interest. Among the pioneering studies of this domain, Martinez et al.\ simulated spontaneous electron-positron pair creation in the Schwinger model on a trapped ion quantum computer \cite{Martinez2016}. Other matter-based platforms such as superconducting qubits \cite{Atas2021,Farrell2024,Klco2018, Cochran2025}, neutral atom \cite{GonzalezCuadra2025,Zhou2022}, trapped-ion system \cite{Nguyen2022} and NMR processors \cite{Singh2025} have also been employed for simulating subatomic physics. Lu et al.\ computed\revnote{Change to ``computed''.} the ground state energy of triton \textsuperscript{3}H, \textsuperscript{3}He, and the alpha particle \textsuperscript{4}He using a quantum frequency processor,demonstrating the applicability of photonic processors to nuclear-physics calculations

Neutral pseudoscalar mesons (K\textsuperscript{0}, B\textsuperscript{0}, B\textsuperscript{0}\textsubscript{s}, D\textsuperscript{0}) play a key role in studies of flavor physics and CP violation in the Standard Model. Weak interactions do not conserve flavour. A neutral meson, which is distinct from its antiparticle, can convert into its antimeson particle and back using weak interaction. CP violation was first observed in the neutral kaon system \cite{Christenson1964}, and K\textsuperscript{0} mixing allowed the charm-quark mass to be estimated before its discovery \cite{GaillardLee1974}. Similarly, the large B\textsuperscript{0} mixing observed by ARGUS indicated a heavy top quark \cite{ARGUS1987,Nierste2025}, and CP violation in the B system established the Kobayashi--Maskawa mechanism as the dominant source of CP violation in quark flavor-changing processes \cite{KM1973,BaBar2001,Belle2001}.

Although unstable particles exhibit non-Hermitian effective dynamics \cite{Bertlmann2006,Caban2005,Benatti1997}, comparatively little attention has been devoted to their quantum simulation\cite{Hu2020,Chen2021}.Hermitian systems like Neutrino flavour oscillations, have been simulated on several matter-based platforms \cite{Arguelles2019,Hall2021,Amitrano2023,Singh2025,NguyenQutrit2023,Turro2025} as well as photonic platform \cite{Marini2014,DiMolfetta2016}. Despite the availability of the theoretical background of open system simulation, only a handful of those systems are simulated due to practical reason \cite{Schlimgen2021,Tischler2018,Suri2023,Barch2026}. A closed system evolves under a unitary propagator, which maps directly onto linear, unitary hardware \cite{Reck1994,clements2016,Carolan2015}. On the contrary, to simulate open-system dynamics on unitary hardware, the non-unitary propagator must first be embedded in a larger unitary through dilation steps such as the Halmos dilation \cite{Halmos1950}, the Sz.-Nagy dilation of contraction semigroups \cite{SzNagy1953,Hu2020}, the Stinespring dilation of quantum channels \cite{Stinespring1955}, the Naimark dilation \cite{Gunther2008,Kivela2024}, the SVD-based diagonal dilation \cite{Schlimgen2022}, the quasi-unitary extension of lossy optical networks \cite{Tischler2018}, the moment-matching dilation \cite{Li2026}, etc.Any contraction can be embedded in a unitary operator on a space of twice the dimension by introducing a single ancilla qubit.\revnote{Suggested replacement: ``Any contraction can be embedded in a unitary operator on a space of twice the dimension by introducing a single ancilla qubit.'' The current sentence lacks a complete grammatical structure.} Among the few reported studies that adopted this approach of non Hermitian quantum system simulation, Sparrow et al.\ simulated a variety of quantum dynamic behaviors in different molecules on a reprogrammable photonic chip \cite{Sparrow2018}; Maraviglia et al.\ simulated coupled PT-symmetric Hamiltonians by programming the dilated propagator onto a reprogrammable chip at each time step \cite{Maraviglia2022}; and Krishna et al.\ emulated the coherent absorption of quantum light on a programmable linear photonic circuit, with the loss channel represented by an ancilla mode \cite{Krishna2026}. To date, within the neutral-meson family, only the kaon has received limited attention. Neutral-kaon mixing, which has Non-Hermitian open quantum system characteristics, has so far been simulated in classical electrical networks through isomorphism \cite{Caruso2016}.Their demonstration uses hand-wired bulk electrical network of lossy components and the observed parameters deviates significantly from those of the kaon-anti kaon system due to parasitic resistances\cite{Caruso2016}.This approach is also specific to only neutral kaon mixing and requires new components for simulating new Hamiltonians. Following this footstep, Chen et al. proposed a simulation of kaon-anti kaon oscillation on an IBM superconducting processor which is yet to appear in a peer reviewed journal \cite{Chen2021}.\revnote{ Bakar: Rewrote this part and eliminated on unpublished journal of kaon mixing. } They reported deviations of hardware results from theory because of device noise, in particular readout (bit-flip) errors, which limits the attainable fidelity.Moreover,their demonstration does not address the oscillation dynamics of the other members of the meson family.To the best of our knowledge, the dynamics of $\Bd$ and $\Bs$ mesons have not yet been simulated on any programmable or fixed platform, either photonic or matter-based.\revnote{ Bakar: Robust claim, need to discuss whether to keep it }

The research gap demands a robust simulation platform that delivers high fidelity and can simulate a whole family of neutral mesons by reprogramming a single device.In this work, we close this gap by simulating the open-system evolution of neutral mesons on a single reprogrammable photonic processor, based on rectangular Clements architecture \cite{clements2016}. We demonstrate it for the $\Kz$, $\Bd$ and $\Bs$ systems on one fixed device, and quantify, through the transfer-matrix fidelity, how faithfully the processor reproduces each of them under realistic, non-uniform losses for different mixing regimes, characterised by x = $\Delta$M/$\Gamma$. The results agree with theoretical predictions.We also uniformly swept the insertion loss of the optical components and calculated fidelity and power transmission characteristics.Finally, we determined the loss resilience of the Photonic Processor by introducing random losses in its components through a Monte Carlo simulation over 100 realizations and measuring the fidelity of the reconstructed unitaries. Thus, we demonstrate the robustness and suitability of this platform for particle physics exploration.

The rest of the paper is organized as follows. First a brief overview of the theoretical background is given in \hyperref[sec:theory]{Section~II}. Then the photonic platform architechture and simulation methods are presented in \hyperref[sec:processor-protocol]{Section~III}. Finally, simulation results are discussed in \hyperref[sec:results]{Section~IV}, followed by the conclusion and acknowledgments.

\section{Theoretical framework}
\label{sec:theory}

\subsection{Effective Hamiltonian of a neutral-meson pair}

We consider a neutral pseudoscalar meson $P^0$ and its antiparticle $\Pbar^0$, with $P^0 =$ $\Kz$, $\Bd$ and $\Bs$.
Weak interactions mix the two flavor states, and both decay. Restricting the dynamics to the
two-dimensional space spanned by $\ket{P^0} \equiv (1,0)^\T$ and $\ket{\Pbar^0} \equiv (0,1)^\T$, the Weisskopf--Wigner
approximation~\cite{Weisskopf1930,Lee1957} gives
\begin{equation}
  i\frac{d}{dt}\begin{pmatrix} c(t)\\ \bar c(t)\end{pmatrix}
  = \Hc \begin{pmatrix} c(t)\\ \bar c(t)\end{pmatrix},
  \qquad \Hc = M - \frac{i}{2}\Gamma,
  \label{eq:schrodinger}
\end{equation}
where $c(t)$ and $\bar c(t)$ are the amplitudes of $\ket{P^0}$ and $\ket{\Pbar^0}$, and $M$ and $\Gamma$ are $2\times 2$ Hermitian matrices,
in addition, $\Gamma \geq 0$. Because $\Hc$ is not Hermitian, the propagator $\Ec(t) = e^{-i\Hc t}$ is not unitary and
the norm of the meson state decreases with time. CPT invariance fixes $\Hc_{11} = \Hc_{22} \equiv m - i\Gamma/2$, so
the only off-diagonal parameters are
\begin{equation}
  \Hc_{12} = M_{12} - \frac{i}{2}\Gamma_{12}, \qquad
  \Hc_{21} = M_{12}^{*} - \frac{i}{2}\Gamma_{12}^{*}.
  \label{eq:offdiag}
\end{equation}
CP violation in mixing requires $M_{12}$ and $\Gamma_{12}$ to carry a relative phase. Throughout this work we
keep that phase, so every case treated below is CP violating.\revnote{The theory part will be unchanged}

The mass eigenstates are
\begin{equation}
  \begin{gathered}
  \ket{P_{1,2}} = p\,\ket{P^0} \pm q\,\ket{\Pbar^0}, \\
  |p|^2 + |q|^2 = 1, \qquad
  \frac{q}{p} = \sqrt{\frac{\Hc_{21}}{\Hc_{12}}},
  \end{gathered}
  \label{eq:eigenstates}
\end{equation}
with complex eigenvalues $E_{1,2} = m_{1,2} - i\Gamma_{1,2}/2 = \Hc_{11} \pm \sqrt{\Hc_{12}\Hc_{21}}$. We label $P_1$ as the lighter state
and define
\begin{equation}
  \begin{gathered}
  \Delta m \equiv m_2 - m_1 > 0, \qquad
  \Delta\Gamma \equiv \Gamma_1 - \Gamma_2, \\
  m = \frac{m_1 + m_2}{2}, \qquad
  \Gamma = \frac{\Gamma_1 + \Gamma_2}{2}.
  \end{gathered}
  \label{eq:definitions}
\end{equation}
For kaons $P_1 = K_S$ and $P_2 = K_L$; for $B_d$ mesons $P_1 = B_L$ and $P_2 = B_H$ etc. The phases of $M_{12}$ and
$\Gamma_{12}$ individually depend on the phase convention of $\ket{P^0}$ and $\ket{\Pbar^0}$; only $\phi \equiv \arg(M_{12}/\Gamma_{12}) = \phi_M - \phi_\Gamma$
is physical~\cite{Branco1999}.

The Particle Data Group~\cite{PDG2026} quotes lifetimes, mass differences and CP asymmetries. Chen \textit{et al.}~\cite{Chen2021} used the leading-order relations $|M_{12}| \simeq \Delta m/2$ and $|\Gamma_{12}| \simeq |\Delta\Gamma|/2$,
which become exact only for $|\cos\phi| = 1$. We instead invert the relations exactly, so that $|q/p|$ fixes
the magnitudes and the phase consistently; this matters little for kaons, where $\phi \simeq \pi$, but changes
$|\Gamma_{12}|$ by several per cent for $\Bd$.\revnote{The theory part will be unchanged}


Four measured quantities thus fix hamiltonian $\Hc$ up to the irrelevant diagonal mass $m$: $\Gamma$, $\Delta m$, $\Delta\Gamma$ and
$|q/p|$. The modulus $|q/p|$ for kaon, for instance, comes from the CP asymmetry in flavor-specific
(semileptonic) decays,
\begin{equation}
  A_L = \frac{1 - |q/p|^2}{1 + |q/p|^2}
  \quad\Longrightarrow\quad
  \left|\frac{q}{p}\right|^2 = \frac{1 - A_L}{1 + A_L}.
  \label{eq:AL}
\end{equation}
In the CP-conserving limit $r = 1$, off-diagonal elements reduce to $M_{12} = -\Delta m/2$ and $\Gamma_{12} = \Delta\Gamma/2$, which are
the inputs used in Sec.~3.1 of Ref.~\cite{Chen2021}.

\subsection{Contructing the dilated propagator}

Since $\Hc_{11} = \Hc_{22}$, the Hamiltonian can be written as
\begin{equation}
  \Hc = \left(m - \frac{i}{2}\Gamma\right)\id + \alpha\,\sigma_x + \beta\,\sigma_y,
  \label{eq:pauli}
\end{equation}
with complex coefficients
\begin{align}
  \alpha &= \frac{\Hc_{12} + \Hc_{21}}{2} = \mathrm{Re}\,M_{12} - \frac{i}{2}\mathrm{Re}\,\Gamma_{12}, \label{eq:alpha}\\
  \beta  &= \frac{i\,(\Hc_{12} - \Hc_{21})}{2} = -\mathrm{Im}\,M_{12} + \frac{i}{2}\mathrm{Im}\,\Gamma_{12}. \label{eq:beta}
\end{align}
The identity part commutes with the rest, so we can factor the propagator as $\Ec(t) = \Ec_1(t)\,\Ec_2(t)\,\Ec_3(t)$
with
\begin{equation}
  \begin{gathered}
  \Ec_1 = e^{-imt}\,\id, \qquad
  \Ec_2 = e^{-\Gamma t/2}\,\id, \\
  \Ec_3 = e^{-i(\alpha\sigma_x + \beta\sigma_y)t}.
  \end{gathered}
  \label{eq:factor}
\end{equation}
$\Ec_1$ is a global phase and drops out of every probability. $\Ec_2$ is the common exponential decay, equal
for both particle antiparticle state. The final term $\Ec_3$ evaluates to be,
\begin{equation}
  \begin{gathered}
  \Ec_3(t) =
  \begin{pmatrix}
    \cos kt & -(\hat\beta + i\hat\alpha)\sin kt\\
    (\hat\beta - i\hat\alpha)\sin kt & \cos kt
  \end{pmatrix},
  \\ \hat\alpha = \frac{\alpha}{k},\ \hat\beta = \frac{\beta}{k}.
  \end{gathered}
  \label{eq:E3}
\end{equation}
Where, $k = \sqrt{\alpha^2 + \beta^2}$. The parameter $k$ is complex whenever $\Delta\Gamma \neq 0$, so $\Ec_3$ itself is not unitary. According to Kraus representation theory~\cite{Kraus1983}, a non-unitary quantum operation can be embedded into a larger unitary operator that acts on the larger system of which the primary system of interest is a part of. Following Ref.~\cite{Chen2021}, we perform the singular-value decomposition
\begin{equation}
  \Ec_3(t) = U(t)\,\mathrm{diag}\{s_1(t), s_2(t)\}\,V^\dagger(t),
  \label{eq:svd}
\end{equation}
where $U$ and $V$ are unitary and $s_{1,2} \geq 0$. Because $\Ec_2$ is proportional to the identity, it can be
absorbed into the diagonal factor:
\begin{equation}
  \begin{gathered}
  \Ec(t) = e^{-imt}\,U(t)\,N(t)\,V^\dagger(t), \\
  N(t) = \mathrm{diag}\{n_1, n_2\}, \qquad
  n_i = s_i\,e^{-\Gamma t/2}.
  \end{gathered}
  \label{eq:N}
\end{equation}
The construction below needs $0 \leq n_i \leq 1$. This follows from the positivity of the decay matrix. Because the eigenvalues of $\Gamma$ are non-negative, $\|\Ec(t)\| \leq 1$ for $t \geq 0$. Hence, the singular values of $\Ec(t)$, namely $n_1$, $n_2$, $\leq 1$ .

Now, we enlarge the total system by introducing ancilla qubit $a$ along with meson qubit $s$. With the ordering $\ket{s}\otimes\ket{a}$, we dilate $N(t)$ into a larger but unitary evolution operator $G(t)$ which acts on the composite system, 
\begin{equation}
\begin{aligned}
  G(t)   &= \begin{pmatrix}
            n_1 & -\sqrt{1-n_1^2} & 0 & 0\\
            \sqrt{1-n_1^2} & n_1 & 0 & 0\\
            0 & 0 & n_2 & -\sqrt{1-n_2^2}\\
            0 & 0 & \sqrt{1-n_2^2} & n_2
          \end{pmatrix}
\end{aligned}
\label{eq:G}
\end{equation}
This is the matrix of Eq.~(18) in Ref.~\cite{Chen2021}, with $a \to n_1$ and $b \to n_2$. The complete two-qubit
evolution is
\begin{equation}
  \Uc(t) = \left(U \otimes \id_a\right)\,G\,\left(V^\dagger \otimes \id_a\right),
  \label{eq:Ufull}
\end{equation}
a $4\times 4$ unitary for every $t$. Taking the projection using $|0_a\rangle\langle 0_a|$ returns the meson's evolution operator,
\begin{equation}
  \bra{0}_a\,\Uc(t)\,\ket{0}_a = U N V^\dagger = e^{imt}\,\Ec(t),
  \label{eq:projection}
\end{equation}
which is $\Ec(t)$ up to a global phase. Similarly, $|1_a\rangle\langle 1_a|$ projector returns the environment's evolution, namely, $\langle 1_a|\Uc |0_a\rangle = U\,\mathrm{diag}\{\sqrt{1-n_1^2}, \sqrt{1-n_2^2}\}\,V^\dagger$.

Projection onto the ancillary state $|1\rangle_a$ gives the amplitude map into the auxiliary decay sector. The summed probability in this sector equals the total decay probability. Individual ancillary outputs are not assigned to specific physical decay channels.

For an initial $\ket{P^0}\otimes\ket{0}_a$ the three measured probabilities are
\begin{align}
  P(P^0 \to P^0) &= |\langle 0_s 0_a|\,\Uc(t)\,|0_s 0_a\rangle|^2 = |g_+(t)|^2, \label{eq:Psurv}\\
  P(P^0 \to \Pbar^0) &= |\langle 1_s 0_a|\,\Uc(t)\,|0_s 0_a\rangle|^2 = \left|\frac{q}{p}\right|^2 |g_-(t)|^2, \label{eq:Posc}\\
  P_{\mathrm{dec}}(t) &= \sum_{j=0,1} |\langle j_s 1_a|\,\Uc(t)\,|0_s 0_a\rangle|^2 \nonumber\\
  &= 1 - P(P^0 \to P^0) - P(P^0 \to \Pbar^0), \label{eq:Pdec}
\end{align}
with $g_\pm(t) = \left(e^{-iE_2 t} \pm e^{-iE_1 t}\right)/2$, so that
\begin{equation}
  |g_\pm(t)|^2 = \frac{1}{4}\left[e^{-\Gamma_1 t} + e^{-\Gamma_2 t} \pm 2e^{-\Gamma t}\cos(\Delta m\,t)\right].
  \label{eq:gpm}
\end{equation}

\section{Photonic processor and simulation method}
\label{sec:processor-protocol}

\subsection{Processor architecture and four-mode encoding}
\label{sec:processor-architecture}

\begin{figure*}[t]
  \centering
  \begin{minipage}[t]{\textwidth}
    \raggedright\textbf{(a)}\par\vspace{2pt}
    \includegraphics[width=\linewidth,trim=5 10 5 10,clip]{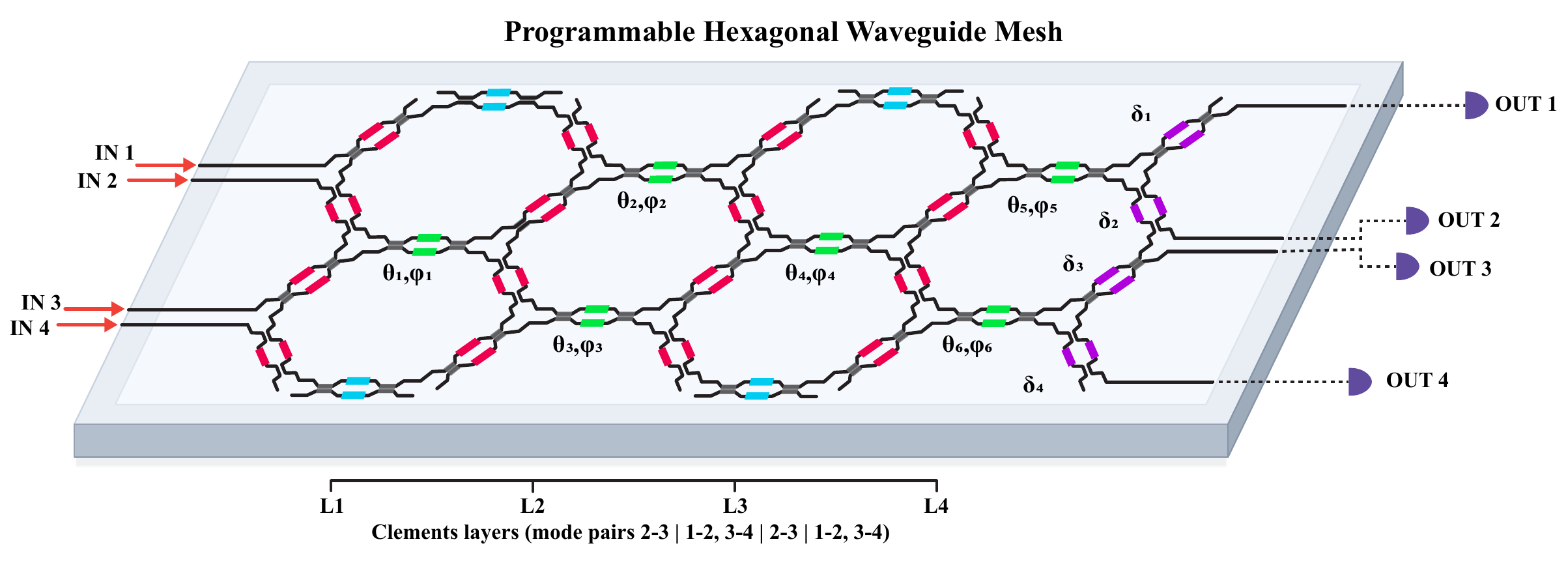}
  \end{minipage}\par\vspace{6pt}
  \begin{minipage}[t]{0.335\textwidth}
    \raggedright\textbf{(b)}\par\vspace{2pt}
    \includegraphics[width=\linewidth]{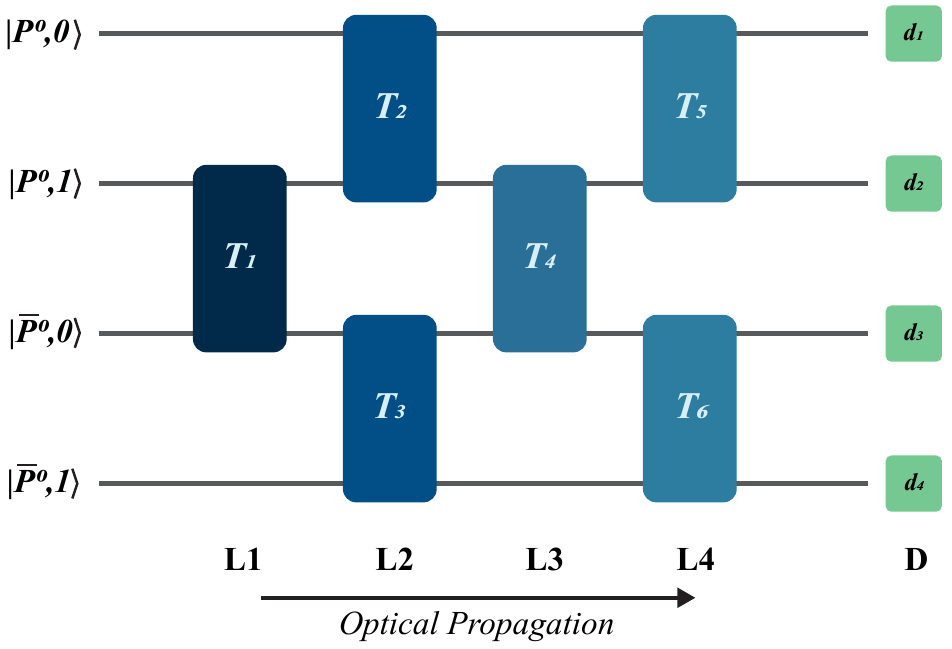}
  \end{minipage}\hfill
  \begin{minipage}[t]{0.412\textwidth}
    \raggedright\textbf{(c)}\par\vspace{2pt}
    \includegraphics[width=\linewidth,trim=8 34 8 34,clip]{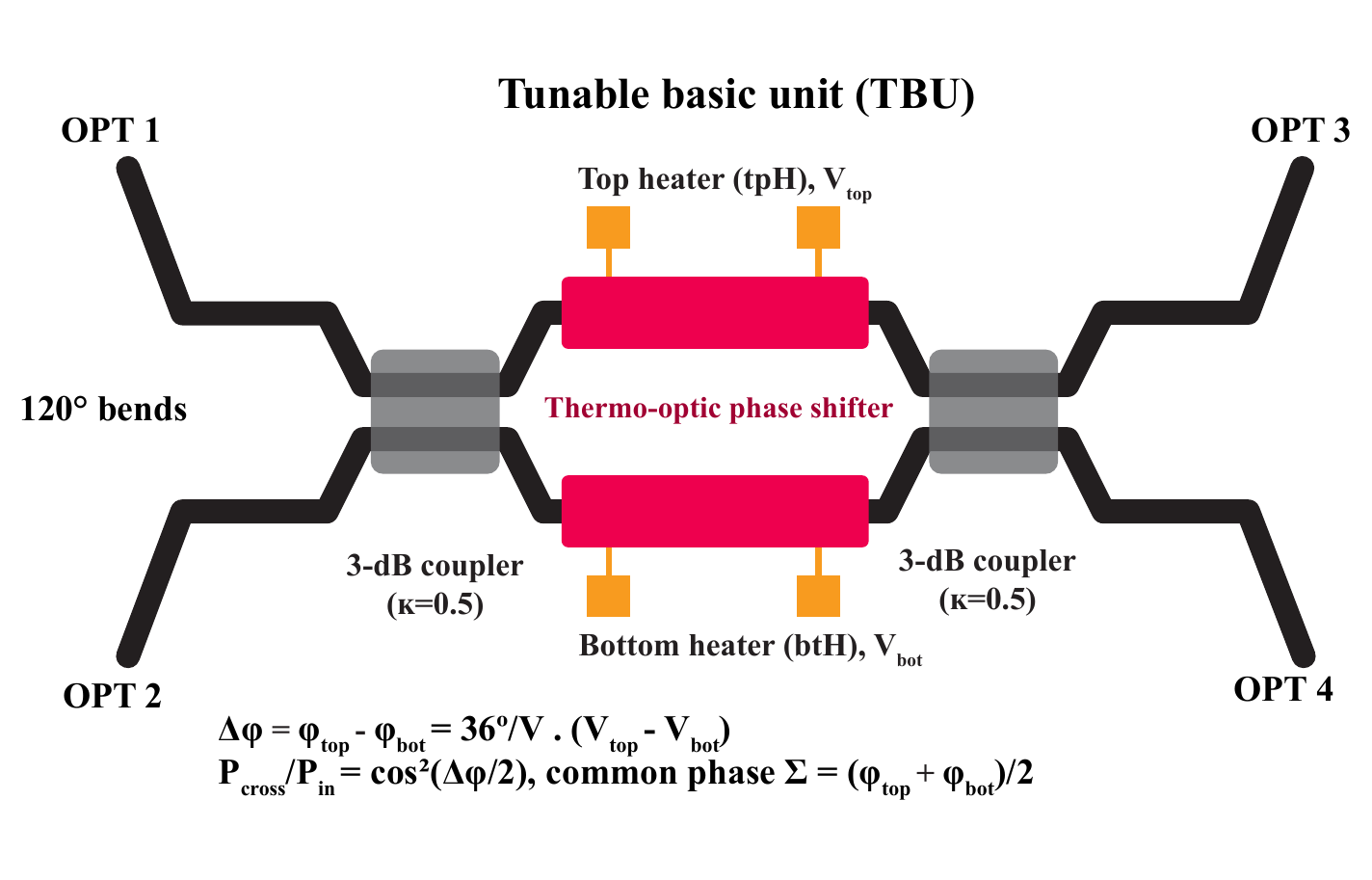}
  \end{minipage}\hfill
  \begin{minipage}[t]{0.223\textwidth}
    \raggedright\textbf{(d)}\par\vspace{2pt}
    \includegraphics[width=\linewidth,trim=10 5 10 4,clip]{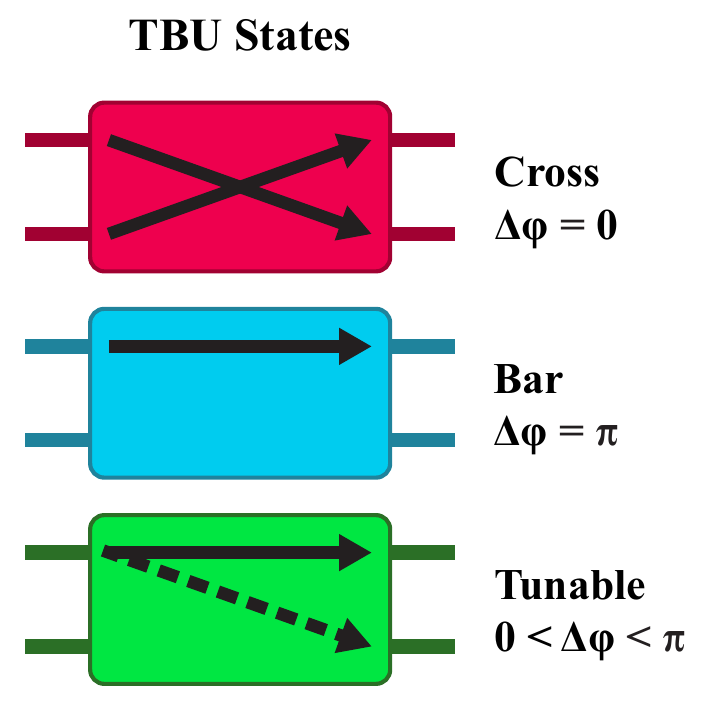}
  \end{minipage}
  \caption{Programmable photonic processor. (a)~Hexagonal waveguide mesh with inputs IN1--IN4 and outputs OUT1--OUT4. The six mixing elements are labeled by their parameters $\theta_\ell,\varphi_\ell$ and the four output trims by $\delta_1,\ldots,\delta_4$; the bracket below the chip marks the Clements layers L1--L4 and the mode pairs that each layer couples. The drawing is schematic and does not reproduce the circuit netlist. (b)~Logical Clements layout of Eq.~\eqref{eq:processor-layout} acting on the basis states of Table~\ref{tab:four-mode-encoding}: six two-mode mixing elements $T_1$--$T_6$ arranged in layers L1--L4, followed by the output phase screen $D$ with phases $d_1,\ldots,d_4$; light propagates from left to right. (c)~Tunable basic unit: a balanced Mach--Zehnder interferometer formed by two 3-dB couplers ($\kappa=0.5$) and a thermo-optic phase shifter on each arm, driven by the voltages $V_{\mathrm{top}}$ and $V_{\mathrm{bot}}$. In the circuit model $\Delta\phi=\phi_{\mathrm{top}}-\phi_{\mathrm{bot}}=(36^\circ/\mathrm{V})(V_{\mathrm{top}}-V_{\mathrm{bot}})$, the cross-port transmission is $P_{\mathrm{cross}}/P_{\mathrm{in}}=\cos^2(\Delta\phi/2)$, and $\Sigma=(\phi_{\mathrm{top}}+\phi_{\mathrm{bot}})/2$ is the common phase. (d)~Cross ($\Delta\phi=0$), bar ($\Delta\phi=\pi$), and tunable ($0<\Delta\phi<\pi$) states.}
  \label{fig:processor-encoding}
\end{figure*}

The processor is a hexagonal waveguide mesh of tunable basic units (TBUs), an architecture that combines programmable coupling with optical routing~\cite{perez2017,rausell2025}. A $4\times4$ transformation is implemented by six programmable two-mode mixing elements together with routing and phase-control units \revnote{A line is removed, we have 30 TBUs in the figure 1 circuit}. Figure~\ref{fig:processor-encoding}(a) shows the mesh with its input and output paths; the drawing conveys the architecture and is not an extracted circuit netlist. The four optical paths encode the ordered basis
\begin{equation}
 \bigl\{|P^0,0\rangle,\ |P^0,1\rangle,\
 |\bar P^0,0\rangle,\ |\bar P^0,1\rangle\bigr\},
 \label{eq:processor-basis}
\end{equation}
where the first label is the flavor and the second the state of the dilation ancilla (Table~\ref{tab:four-mode-encoding}).

\begin{table}[b]
\caption{Four-mode encoding of the meson--ancilla system. Mode $j$ enters the mesh at input IN$j$ and leaves it at output OUT$j$.}
\label{tab:four-mode-encoding}
\begin{ruledtabular}
\begin{tabular}{cccc}
Mode 1 & Mode 2 & Mode 3 & Mode 4\\
\hline
$|P^0\rangle\otimes|0\rangle_a$ & $|P^0\rangle\otimes|1\rangle_a$ & $|\bar{P}^0\rangle\otimes|0\rangle_a$ & $|\bar{P}^0\rangle\otimes|1\rangle_a$\\
Survives as $P^0$ & Decay sector & Survives as $\bar{P}^0$ & Decay sector\\
\end{tabular}
\end{ruledtabular}
\end{table}

Light injected into IN1 prepares the initial state $|P^0\rangle\otimes|0\rangle_a$. The outputs OUT1 and OUT3 correspond to the two undecayed flavors, while the sum of the normalized powers at OUT2 and OUT4 gives the decay probability,
\begin{equation}
 P_{\mathrm{dec}}(t)=p_{21}(t)+p_{41}(t),
 \label{eq:processor-pdec}
\end{equation}
with the normalized outputs $p_{j1}$ defined in Eq.~\eqref{eq:processor-normalisation} below. Injection into IN3 would prepare the antiparticle with the same initial ancilla state, whereas injection into IN2 or IN4, or into any combination involving those inputs, would not represent an undecayed meson. The two-dimensional meson block retains the non-unitary meson amplitudes [Eq.~\eqref{eq:projection}], while the four-dimensional enlarged space conserves the total norm. In the ideal lossless optical model, the probability is only redistributed among the four outputs, so that $P(\Pz\to\Pz)+P(\Pz\to\Pzb)+P_{\mathrm{dec}}=1$. Measuring the outputs separately therefore reveals whether the meson has survived and, if it has, its flavor. Both decay outputs, OUT2 and OUT4, enter the probability budget rather than being discarded, and optical loss is treated separately from the encoded decay (Sec.~\ref{sec:processor-acquisition}).

\subsection{TBU operation and numerical programming}
\label{sec:processor-programming}

Each TBU is a balanced Mach--Zehnder interferometer (MZI) comprising a 3-dB coupler, followed by a pair of phase shifters that set the arm phases $\phi_{\rm top}$ and $\phi_{\rm bot}$ independently, and finally a second 3-dB coupler [Fig.~\ref{fig:processor-encoding}(c)]. The arm phases can be set with thermo-optic or electro-optic phase shifters. With $\Delta\phi=\phi_{\rm top}-\phi_{\rm bot}$ and $\Sigma=(\phi_{\rm top}+\phi_{\rm bot})/2$, the ideal field transfer matrix is
\begin{equation}
 B(\Delta\phi,\Sigma)=i e^{i\Sigma}
 \begin{pmatrix}
  \sin(\Delta\phi/2)&\cos(\Delta\phi/2)\\
  \cos(\Delta\phi/2)&-\sin(\Delta\phi/2)
 \end{pmatrix}.
 \label{eq:processor-tbu}
\end{equation}
The cross-port power fraction is therefore $\cos^2(\Delta\phi/2)$: $\Delta\phi=0$ gives the cross state, $\Delta\phi=\pi$ the bar state, and intermediate values a tunable splitter [Fig.~\ref{fig:processor-encoding}(d)]. The common phase $\Sigma$ becomes relevant once the TBU is embedded in an interferometer and is included in the phase bookkeeping.

At each simulated time $t_i$, the dilated target $\mathcal{U}_i$ of Eq.~\eqref{eq:Ufull} is fitted to the rectangular layout of Clements \textit{et al.}~\cite{clements2016},
\begin{equation}
 \begin{aligned}
 \mathcal{U}_i\simeq\widehat{\mathcal{U}}_i
 &=D\,T_6^{(34)}T_5^{(12)}T_4^{(23)}
       T_3^{(34)}T_2^{(12)}T_1^{(23)},\\
 D&=\operatorname{diag}(e^{id_1},\ldots,e^{id_4}),
 \end{aligned}
 \label{eq:processor-layout}
\end{equation}
where the superscripts identify the two coupled modes (numbered from 1), the rightmost transformation acts first, and
\begin{equation}
 T(\theta,\varphi)=
 \begin{pmatrix}
  e^{i\varphi}\cos\theta&-\sin\theta\\
  e^{i\varphi}\sin\theta&\cos\theta
 \end{pmatrix}.
 \label{eq:processor-logical-t}
\end{equation}
In each elementary transformation, $\theta$ sets the splitting ratio and $\varphi$ the relative phase, and $D$ contains the four output phases [Fig.~\ref{fig:processor-encoding}(b)]. Six such transformations, the number required for a universal four-mode interferometer, act on nearest-neighbor modes. In the analytical procedure of Ref.~\cite{clements2016}, multiplication from the left by $T$ mixes neighboring rows and multiplication from the right mixes neighboring columns; alternating the two nulls successive diagonals of the target without restoring previously nulled elements, the remaining triangular matrix is diagonal because it is unitary, and rearranging the operations yields Eq.~\eqref{eq:processor-layout}. Here the parameters $\{\theta_\ell,\varphi_\ell,d_j\}$ are instead obtained by bounded nonlinear least-squares fitting with up to 200 random initializations. The conversion of the fitted parameters into physical arm phases is closed form and includes the phase contributions of the bypass units and of the four output trims. \newchange{Specifically, the logical mixing matrix is factorized as

\begin{equation}
\begin{aligned}
T(\theta,\varphi)&=R(\theta)\operatorname{diag}(e^{i\varphi},-1),\\
R(\theta)&=\begin{pmatrix}
\cos\theta&\sin\theta\\
\sin\theta&-\cos\theta
\end{pmatrix}.
\end{aligned}
\end{equation}

For $\Delta\phi=\pi-2\theta$, Eq.~\eqref{eq:processor-tbu} gives
\begin{equation}
\begin{aligned}
B(\pi-2\theta,\Sigma)&=i e^{i\Sigma}R(\theta),\\
T(\theta,\varphi)&=B(\pi-2\theta,\Sigma)\,Q,\\
Q&=e^{-i(\Sigma+\pi/2)}\operatorname{diag}(e^{i\varphi},-1).
\end{aligned}
\end{equation}
The corresponding arm phases are $\phi_{\rm top}=\Sigma+\pi/2-\theta$ and $\phi_{\rm bot}=\Sigma-\pi/2+\theta$, modulo $2\pi$. The diagonal phase factors are implemented by routing TBUs distributed throughout the mesh, whose settings also compensate for the common phases introduced by the mixing and bypass units. The four final routing TBUs implement the output phase screen $D$. These four phases are generally different and  $D$ is not a single global phase.}

The protocol requires only control of the optical phase and does not depend on a specific actuator. The circuit model assumes the ideal calibration $\phi(V)=2\pi V/(10\,\mathrm{V})$, equivalent to $36^\circ/\mathrm{V}$; arm phases are wrapped modulo $2\pi$ before conversion to voltages in $[0,10)\,\mathrm{V}$. 

\subsection{Circuit model and independently programmed instants}
\label{sec:processor-model}

The mesh is simulated with Ansys Lumerical INTERCONNECT at a single optical carrier of $193.1\,\mathrm{THz}$ (wavelength $\approx1552.5\,\mathrm{nm}$). Each TBU is modeled with its two couplers, phase shifters on both arms, and the bends that connect it to its neighbors. Table~\ref{tab:processor-model} lists the geometry and the loss assumptions; the loss values follow the experimental TBU characterization of Ref.~\cite{perez2017} (Supplementary Note 2 and Supplementary Table 1 therein).

\begin{table}[b]
\caption{Circuit-model settings. For the three stochastic entries, $\mu$ is the mean and $a$ the spread used by the generator; the Gaussian standard deviation is taken as $a/3$, and negative draws are resampled.}
\label{tab:processor-model}
\begin{ruledtabular}
\begin{tabular}{@{}p{0.50\columnwidth}l@{}}
\raggedright Quantity & Setting\\
\hline
\raggedright Optical carrier & $193.1\,\mathrm{THz}$\\
\raggedright TBU path length & $975\,\mu\mathrm{m}$\\
\raggedright Bend radius & $10\,\mu\mathrm{m}$\\
\raggedright Phase-shifter arm (path after bend arcs) & $901.70\,\mu\mathrm{m}$\\
\raggedright Propagation loss (one draw per TBU) & $\mu=2.40$, $a=0.30$~dB/cm\\
\raggedright Coupler insertion loss (one draw per coupler) & $\mu=0.15$, $a=0.10$~dB\\
\raggedright Excess bend loss (total per TBU path) & $\mu=0.10$, $a=0.02$~dB\\
\raggedright Arm phase range & $0$ to $2\pi$ (linear, $0$--$10$~V)\\
\end{tabular}
\end{ruledtabular}
\end{table}

Losses are nonnegative Gaussian random variables assigned independently to each TBU. Within one TBU, both arms share the propagation-loss parameter and the two 3-dB couplers receive independent insertion losses. The excess bend loss is a total for one path through the TBU and is distributed among the bend elements in proportion to their angles; propagation through the bend arcs is included separately. At the nominal values the loss budget is $2(0.15)+2.40(0.0975)+0.10=0.634\,\mathrm{dB}$ per TBU path. One loss realization is drawn and kept fixed for the simulations of Sec.~\ref{sec:res-dynamics}, as it would be on a fabricated chip.

All three target generators retain CP violation in mixing using the inputs in Table~\ref{tab:params}. In every case the positivity of the decay matrix ensures that the propagator is a contraction and that the singular values entering the dilation lie between zero and one.

\cpsuggestion{4}{\textbf{Ctrl+F: ``CP-conserving mixing''.} Keep this paragraph for the uploaded generator. If $B_d$ is rerun with CP violation, replace the approximation statement with: ``All three target generators retain CP violation in mixing using the inputs in Table~\ref{tab:params}.'' Compute the Hamiltonian from those inputs using exact inversion; do not merely change their declarations or hard-code rounded derived values.}

\begin{figure*}[t]
 \centering
 \includegraphics[width=\textwidth]{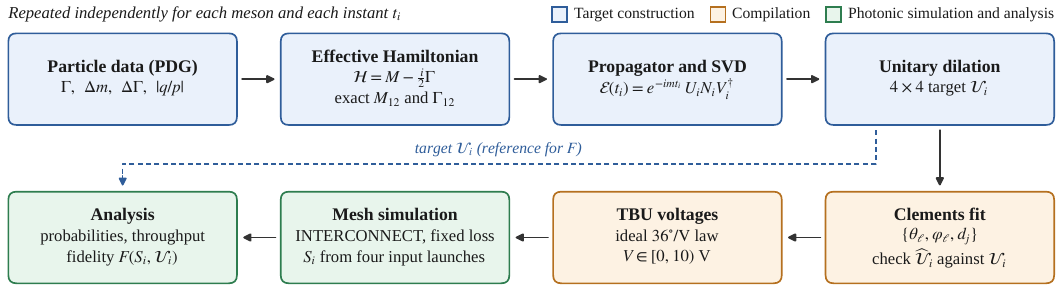}
 \caption{Simulation protocol, repeated independently for each instant $t_i$ and each meson. Particle-data inputs fix the effective Hamiltonian [Eqs.~\eqref{eq:schrodinger}--\eqref{eq:definitions}]; its propagator is singular-value decomposed and dilated into the four-mode target $\mathcal{U}_i$ [Eqs.~\eqref{eq:svd}--\eqref{eq:Ufull}], which is fitted to the Clements layout of Eq.~\eqref{eq:processor-layout}. The fitted parameters are converted into TBU voltages and loaded into the hexagonal mesh, simulated in INTERCONNECT with a fixed loss realization. The complex transfer matrix $S_i$, obtained from four sequential input launches, yields the meson probabilities, the throughput, and \newchange{the fidelity $F(S_i,\widehat{\mathcal{U}}_i)$, with the compiled matrix supplied by the Clements-fit block. }}
 \label{fig:processor-workflow}
\end{figure*}

Each meson system is evolved over $0\leq\Gamma t<4$ on $N_t=500$ instants,
\begin{equation}
 t_i=\frac{4i}{\Gamma N_t},\qquad i=0,\ldots,N_t-1,
 \label{eq:processor-grid}
\end{equation}
which provides about 29 samples per $B_s$ oscillation period for the parameters of Sec.~\ref{sec:res-regimes}. At every instant the target is reconstructed for the same initial state and loaded as a separate, static optical transformation. Successive $\mathcal{U}_i$ are not cascaded: in general the chosen dilations need not compose as a time-evolution semigroup on the enlarged space. Figure~\ref{fig:processor-workflow} summarizes the procedure.

\subsection{Transfer-matrix acquisition, normalization, and fidelity}
\label{sec:processor-acquisition}

For each programmed setting, an optical network analyzer (ONA) element~\cite{ansysONA} launches light into each of the four inputs in turn and records the complex transmission to all four outputs. This gives the transfer matrix $S_{jk}$, with rows denoting outputs and columns inputs. Because the ONA returns complex field amplitudes~\cite{ansysONA}, $|S_{jk}|^2$ is the power transmission. For input IN1 we define the collected throughput $\eta_1$ and the normalized outputs
\begin{equation}
 \eta_1(t_i)=\sum_{k=1}^{4}|S_{k1}(t_i)|^2,\qquad
 p_{j1}(t_i)=\frac{|S_{j1}(t_i)|^2}{\eta_1(t_i)}.
 \label{eq:processor-normalisation}
\end{equation}
The flavor probabilities are inferred from $p_{11}$ and $p_{31}$, and the decay probability from $p_{21}+p_{41}$ [Eq.~\eqref{eq:processor-pdec}]. All four paths enter the denominator, so there is no selection on the ancilla outcome. The normalization does condition on optical collection; however, it cancels a common attenuation factor but does not undo path-dependent loss. Optical loss is distinct from the encoded meson decay, and the throughput, $-10\log_{10}\eta_1$ in decibels, is therefore reported separately (Sec.~\ref{sec:res-uniform}).

The quality of each programmed transformation is quantified by the fidelity~\cite{clements2016}
\begin{equation}
 F(S,U)=\frac{|\operatorname{tr}(U^\dagger S)|^2}
 {4\operatorname{tr}(S^\dagger S)}
 \label{eq:processor-fidelity}
\end{equation}
between the simulated transfer matrix $S$ and a unitary reference $U$. This loss-normalized overlap is invariant under a global phase and, by the Cauchy--Schwarz inequality, equals unity if and only if $S=cU$ for a nonzero scalar $c$. It is therefore insensitive to uniform loss and is lowered only by imperfections that distort the transformation, such as unbalanced loss or errors in splitting ratios and phases.\newchange{All fidelities in Sec. IV are computed between the full simulated transfer matrix \(S_i\) and the compiled unitary \(\widehat{\mathcal U}_i\) reconstructed from the fitted Clements parameters, \newchange{using \(F(S_i,\widehat{\mathcal U}_i)\) in Eq.~\eqref{eq:processor-fidelity}.} This measures the accuracy with which the photonic circuit implements the compiled transformation. }


The circuit calculations use classical complex fields. For a linear passive network the same matrix gives the single-photon transition probabilities $|S_{jk}|^2$, and for equally efficient detectors the conditional detection frequencies follow the normalization of Eq.~\eqref{eq:processor-normalisation}. The simulated intensities therefore correspond to a single-excitation encoding, in which one meson is represented by one photon and its flavor and survival by the photon's path.

\section{Results and Discussion}
\label{sec:results}

\subsection{Meson dynamics on one programmed mesh}
\label{sec:res-regimes}
\label{sec:res-dynamics}

\begin{table*}[t]
\caption{Input parameters and derived effective-Hamiltonian quantities for the three simulated systems. The inputs are particle-data values~\cite{PDG2026,HFLAV2026}. Here $x=\Delta m/\Gamma$, $y=\Delta\Gamma/2\Gamma$ with $\Delta\Gamma=\Gamma_1-\Gamma_2$, and $\phi=\arg(M_{12}/\Gamma_{12})$ with $\Gamma_{12}$ chosen real and positive. All systems are evolved over $0\leq\Gamma t<4$ with $N_t=500$ instants.}
\label{tab:params}
\begin{ruledtabular}
\begin{tabular}{lccc}
 & $\Kz$--$\Kzb$ & $\Bd$--$\Bdb$ & $\Bs$--$\Bsb$\\
\hline
Lifetime input & $\tau_S=0.8954\times10^{-10}$~s, $\tau_L=5.116\times10^{-8}$~s & $\tau=1.5113$~ps & $\tau=1.515$~ps\\
$\Delta m$ ($\mathrm{s}^{-1}$) & $0.5293\times10^{10}$ & $0.5068\times10^{12}$ & $17.766\times10^{12}$\\
Width difference & $\Gamma_S-\Gamma_L$ (from $\tau_S$, $\tau_L$) & $\Delta\Gamma_d/\Gamma_d=0.005$ & $\Delta\Gamma_s/\Gamma_s=0.118$\\
CP input & $A_L=3.32\times10^{-3}$ & $|q/p|=1.0010$ & $(1-|q/p|^2)/4=-1.5\times10^{-4}$\\[3pt]
$\Gamma$ ($\mathrm{s}^{-1}$) & $5.594\times10^{9}$ & $6.617\times10^{11}$ & $6.601\times10^{11}$\\
$x=\Delta m/\Gamma$ & $0.946$ & $0.766$ & $26.9$\\
$y=\Delta\Gamma/2\Gamma$ & $0.997$ & $0.0025$ & $0.059$\\
$|q/p|$ & $0.99669$ & $1.00100$ & $1.00030$\\
$|M_{12}|$ ($\mathrm{s}^{-1}$) & $2.647\times10^{9}$ & $2.534\times10^{11}$ & $8.883\times10^{12}$\\
$\Gamma_{12}$ ($\mathrm{s}^{-1}$) & $5.574\times10^{9}$ & $1.730\times10^{9}$ & $3.931\times10^{10}$\\
$\phi$ & $180.4^\circ$ & $163.0^\circ$ & $172.2^\circ$\\
\end{tabular}
\end{ruledtabular}
\end{table*}

\begin{figure*}[p]
  \centering
  \setlength{\unitlength}{0.94\textwidth}
  \includegraphics[width=0.94\textwidth]{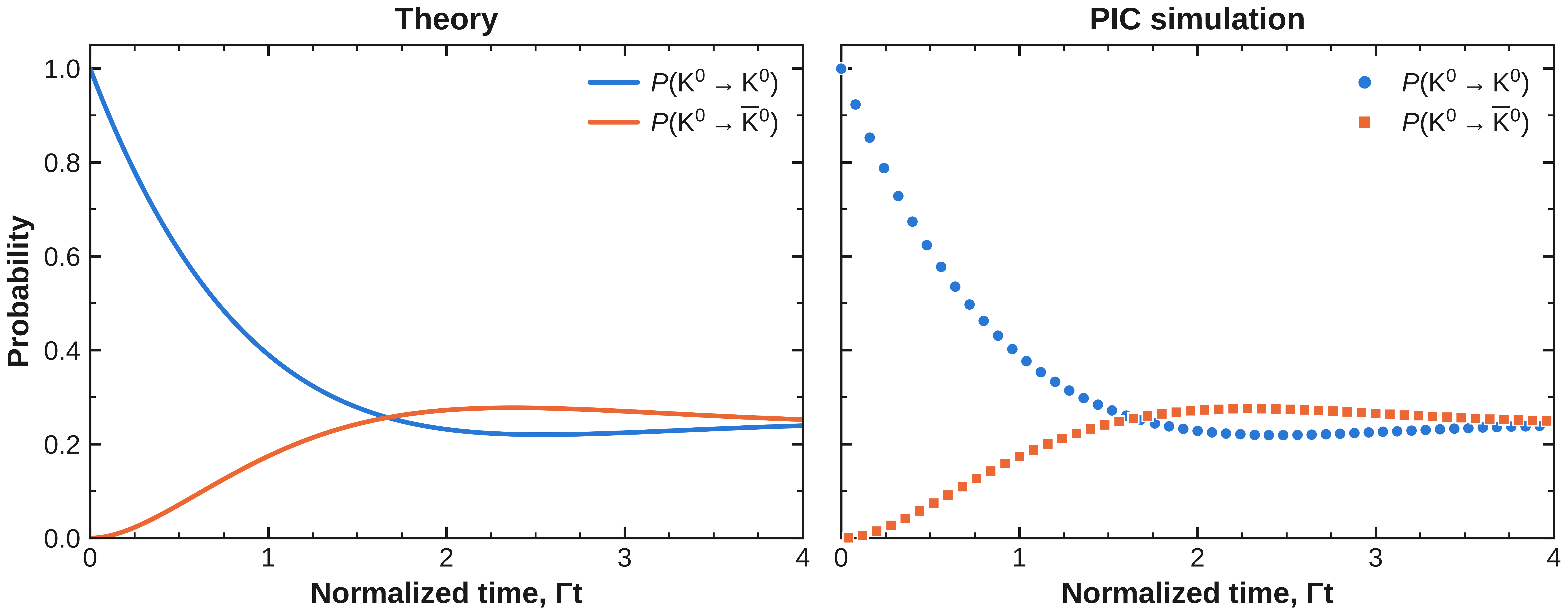}%
  \begin{picture}(0,0)\put(-1.0,0.372){\textbf{(a)}}\end{picture}\\[1mm]
  \includegraphics[width=0.94\textwidth]{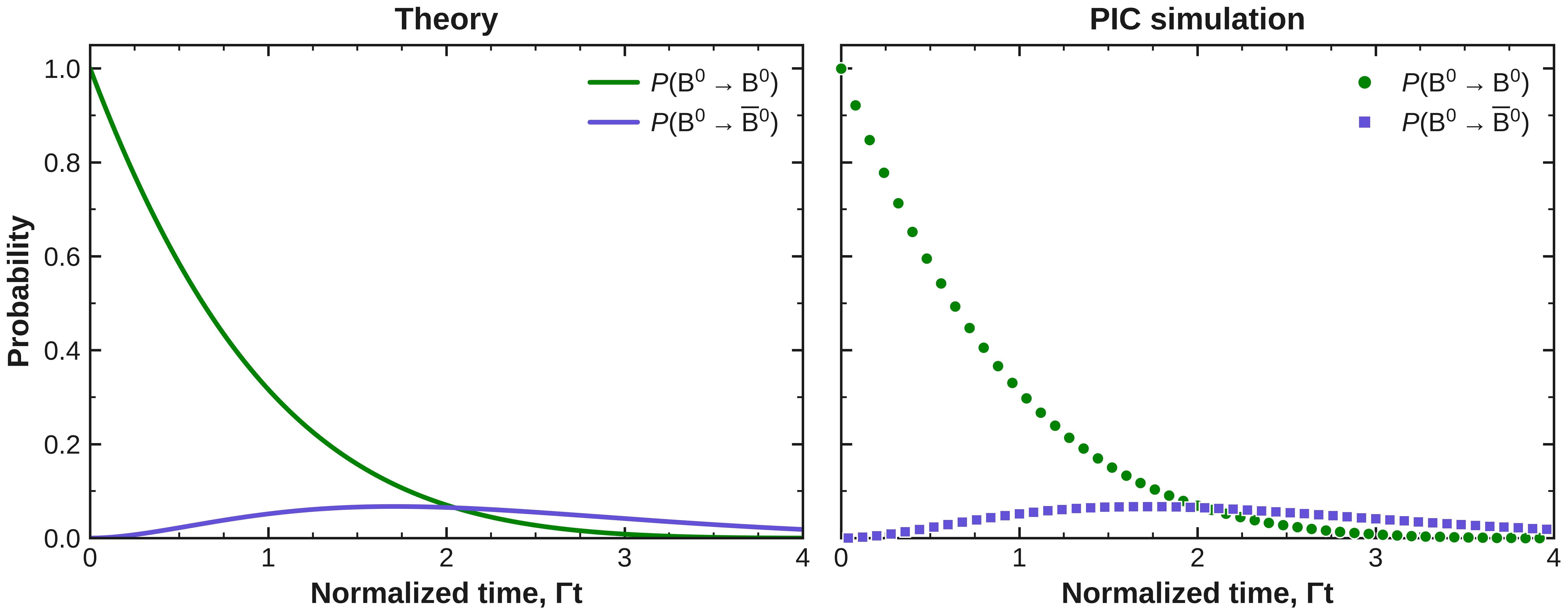}%
  \begin{picture}(0,0)\put(-1.0,0.372){\textbf{(b)}}\end{picture}\\[1mm]
  \includegraphics[width=0.94\textwidth]{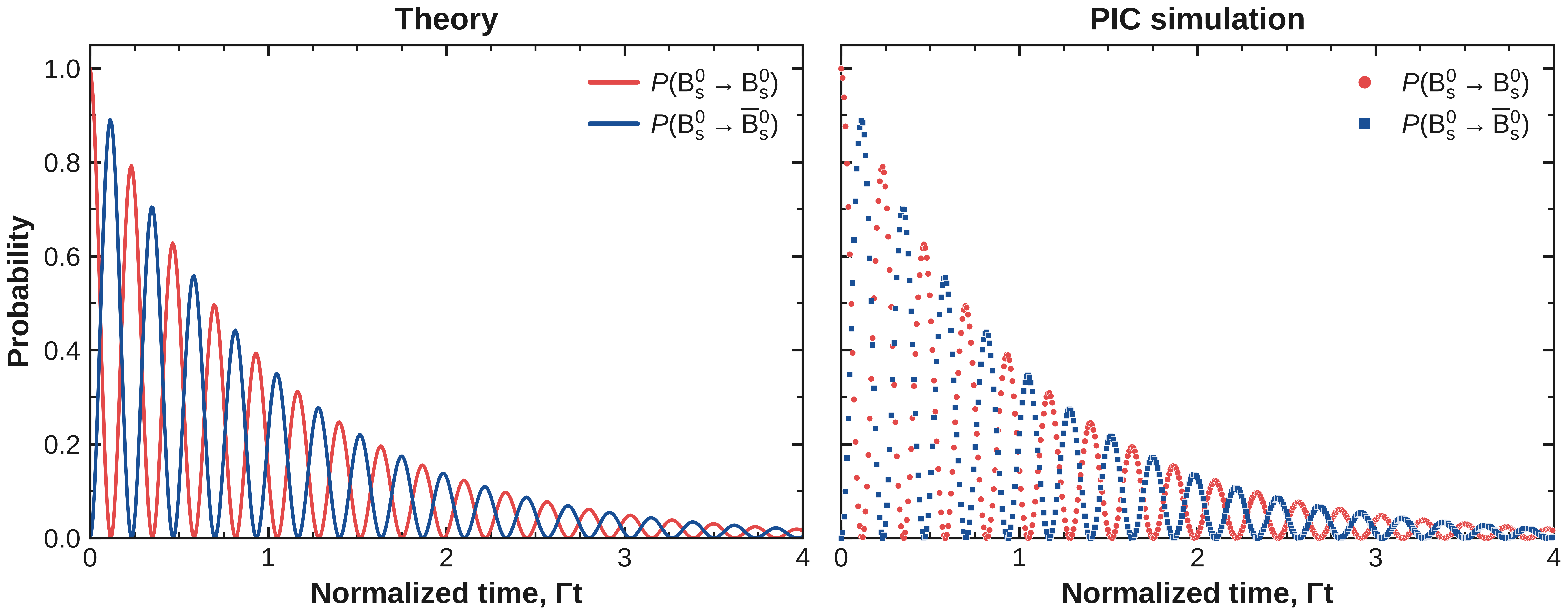}%
  \begin{picture}(0,0)\put(-1.0,0.372){\textbf{(c)}}\end{picture}
  \caption{Survival probability $P(\Pz\to\Pz)$ and oscillation probability $P(\Pz\to\Pzb)$ of an initial $\Pz$ versus normalized time $\Gamma t$ for (a)~$\Kz$--$\Kzb$, (b)~$\Bd$--$\Bdb$, and (c)~$\Bs$--$\Bsb$, with the parameters of Table~\ref{tab:params}. Left: theory, Eqs.~\eqref{eq:Psurv}, \eqref{eq:Posc}, and \eqref{eq:gpm}. Right: simulation of the lossy hexagonal mesh, given by the normalized outputs $p_{11}$ (circles) and $p_{31}$ (squares) of Eq.~\eqref{eq:processor-normalisation}. Each instant is an independently programmed $4\times4$ unitary; for clarity, every tenth of the 500 instants is plotted in (a) and (b), and all instants in (c). The three systems run on the same device, with one fixed loss realization, and differ only in the programmed TBU voltages.}
  \label{fig:prob}
\end{figure*}

Table~\ref{tab:params} lists the input parameters of the three systems and the effective-Hamiltonian quantities reconstructed from them by inverting Eqs.~\eqref{eq:eigenstates} and \eqref{eq:definitions} exactly. The kaon, $B_d$, and $B_s$ targets retain CP violation in mixing, with $\phi=180.4^\circ$, $163.0^\circ$ and $172.2^\circ$ respectively.Because $\Gamma$ is absorbed into the time variable, the physical time scales, nanoseconds for kaons and picoseconds for $B$ mesons, never enter the hardware; only $x$, $y$, and $|q/p|$ are programmed. Figure~\ref{fig:prob} compares the analytic probabilities with the normalized outputs $p_{11}$ and $p_{31}$ of Eq.~\eqref{eq:processor-normalisation}, obtained from the INTERCONNECT model of the mesh. For each meson the processor was reprogrammed at the $N_t=500$ instants of Eq.~\eqref{eq:processor-grid}, and all 1500 configurations run on the same circuit with the same loss realization; only the TBU voltages differ. With classical light these normalized intensities equal the single-photon transition probabilities (Sec.~\ref{sec:processor-acquisition}).

\cpsuggestion{7}{\textbf{Figure 3.} Check which generator produced the saved $B_d$ curves. If it was the uploaded CP-conserving version and those results are retained, keep the figure. If CP-violating targets are adopted, regenerate the targets, programmed voltages, and affected circuit results before updating Fig.~3(b). The artwork is unchanged here.}

For kaons, the $K_S$ component decays too quickly, even before a complete oscillation can take place. After $\Gamma t=2$, the only mass eigenstate that remains predominantly is $K_L$. As a result, flavor eigenstates $\Kz$ and $\Kzb$ becomes almost equal[Fig.~\ref{fig:prob}(a)]. For $B_d$ mesons, $\Delta\Gamma\ll\Gamma$ and the decay becomes almost independent of flavor. This prevents the flavor states from persisting at a steady nonzero probability, unlike in the kaon system. The $B_s$ system [Fig.~\ref{fig:prob}(c)] completes $x/2\pi\approx4.3$ oscillations per mean lifetime, about 17 within the window, a sharp contrast from the evolution of both kaon and $B_d$.

The fidelity of Eq.~\eqref{eq:processor-fidelity} evaluated between the simulated transfer matrix and the compiled unitary, quantifies the accuracy of the optical implementation. Over the 500 instants, $\langle F\rangle=0.999790\pm0.000047$ for kaons, $0.999815\pm0.000059$ for $B_d$, and $0.999807\pm0.000044$ for $B_s$ (mean and standard deviation over instants), and no instant falls below $F=0.9996$. Because $F$ is insensitive to a common scale factor, these infidelities of $(1.85$--$2.10)\times10^{-4}$ measure the imbalance of the loss among the optical paths of the chip, and their scatter from instant to instant reflects how different unitaries sample one fixed loss pattern rather than any additional randomness.The three systems evolve very differently, yet their infidelities are nearly the same. This anticipates the result of Sec.~\ref{sec:res-mc}, which shows that, to leading order, the loss-induced infidelity is independent of the programmed unitary.

\cpsuggestion{9}{\textbf{Fidelity and Monte Carlo results.} Retain $0.999815$ and the $B_d$ Table IV statistics only if retaining the targets that generated them. A CP-violating rerun requires recalculating the affected circuit fidelities, Monte Carlo statistics, and fitted coefficient. A small probability change does not guarantee unchanged fidelity. The uniform-loss argument and leading-order topology-dependent coefficient remain applicable.}

The construction of Sec.~\ref{sec:theory} maps any CPT-symmetric two-state Hamiltonian with a positive decay matrix onto a $4\times4$ unitary, and a hexagonal mesh can host a universal interferometer of this size~\cite{clements2016,rausell2025}, so any decaying two-level system described by such a Hamiltonian, are accessible by reprogramming alone.In every case the flavor and the ancilla occupy distinct waveguides, so the four outputs, including the decay sector, are recorded simultaneously and without postselection. In the superconducting-qubit simulation of kaon mixing~\cite{Chen2021}, readout bit flips, at an average readout error of about $2\%$, exchanged system and environment outcomes; the waveguide encoding has no direct analog of this error.

\subsection{Uniform insertion loss and path balance}
\label{sec:res-uniform}

To investigate the effects of uniform loss, we assigned the same insertion loss $L$ to every TBU, and swept $L$ from $0$ to $5$~dB in steps of $0.5$~dB for the neutral meson evolution with fixed programmed voltages. At each $L$ we recorded the fidelity averaged over the programmed instants and the total power transmission of each input mode $j$,
\begin{equation}
  T_j=\sum_{i=1}^{4}|S_{ij}|^2 .
  \label{eq:Tj}
\end{equation}

\begin{figure*}[t]
  \centering
  \includegraphics[width=0.8\textwidth]{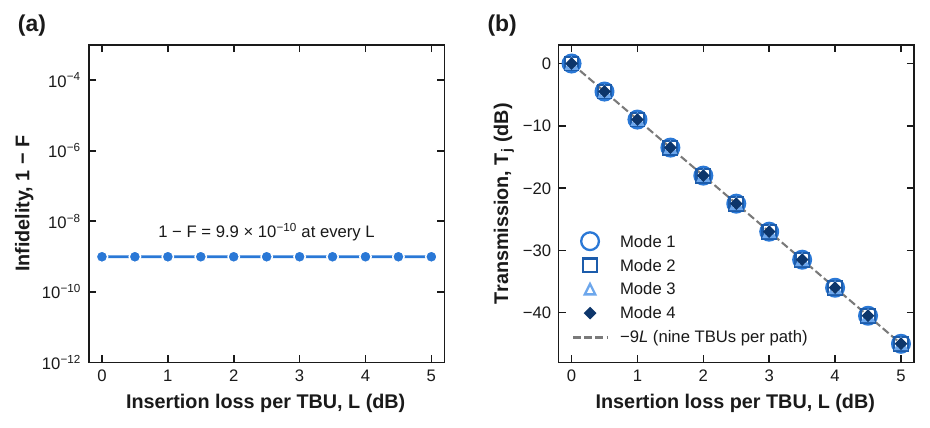}
  \caption{Uniform insertion loss: the same loss $L$ is applied to every TBU, split equally between its two 3-dB couplers, with all other elements lossless, for the $\Kz$--$\Kzb$ evolution with fixed programmed voltages. (a)~Infidelity $1-F$ [Eq.~\eqref{eq:processor-fidelity}], averaged over the programmed instants, versus $L$ on a logarithmic axis; \newchange{the constant level of $9.9\times10^{-10}$ is the loss-independent baseline infidelity.} (b)~Total power transmission $T_j$ of input modes $j=1$--4 [Eq.~\eqref{eq:Tj}], in decibels, versus $L$. The markers of the four modes coincide; the dashed line is $T_j=-9L$, the attenuation of the nine TBUs that every input-output path crosses.}
  \label{fig:uniform}
\end{figure*}

The infidelity does not change with \(L\), it remains at a baseline value of \(9.9\times10^{-10}\), five orders of magnitude below the loss-limited value of Sec.~\ref{sec:res-dynamics} [Fig.~\ref{fig:uniform}(a)].  The transmissions of the four modes coincide and follow $T_j=-9L$ in decibels, with a fitted slope of $-9.000$ and an intercept below $10^{-13}$~dB [Fig.~\ref{fig:uniform}(b)].For all the programmed unitaries, every input-output path crosses exactly nine TBUs. A uniform field attenuation $\eta=10^{-L/20}$ per TBU then gives
\begin{equation}
  S=\eta^{9}\,\widehat{\mathcal{U}} ,\qquad
  T_j=\eta^{18}\sum_{i}|\widehat{\mathcal{U}}_{ij}|^{2}=\eta^{18},
  \label{eq:uniformS}
\end{equation}
because the columns of a unitary matrix have unit norm. The fidelity is invariant under $S\to cS$, and the normalization of Eq.~\eqref{eq:processor-normalisation} divides out $\eta^{18}$, neither $F$ nor the inferred probabilities change, and the transmission is the same for every programmed unitary and every meson.Uniform loss therefore costs only throughput. At the nominal budget of $0.634$~dB per TBU (Table~\ref{tab:processor-model}), the insertion loss is $5.7$~dB and $27\%$ of the injected power reaches the detectors. In a single-photon experiment this lengthens the acquisition time needed for a given precision by the factor $10^{0.9L}$ but does not bias the simulated dynamics, and it leaves the unit-to-unit variation of the loss as the only loss-related source of error.

\subsection{Nonuniform loss and fabrication tolerance}
\label{sec:res-mc}

\begin{table*}[t]
\caption{Monte Carlo statistics of the chip fidelity $F_c$ over 100 random-loss chips per loss spread $a=3\sigma$ and per meson; the same 100 loss draws, rescaled by $a$, are used throughout. Here $s$ is the standard deviation over chips, $P_{10}$ the 10th percentile, and $Y_{0.999}$ the fraction of chips with $F_c\geq0.999$. The last column gives the coefficient $\kappa$ of the fit $1-\langle F\rangle=\kappa a^2$ [Eq.~\eqref{eq:kappa}] for each meson.}
\label{tab:mc}
\begin{ruledtabular}
\begin{tabular}{llccccccc}
Meson & $a$ (dB) & $\langle F\rangle$ & $s$ & Median & $P_{10}$ & $\min F_c$ & $Y_{0.999}$ & $\kappa$ ($10^{-3}$~dB$^{-2}$)\\
\hline
$\Kz$ & 0.1 & 0.999911 & $5.6\times10^{-5}$ & 0.999930 & 0.999830 & 0.999737 & 1.00 & 8.86\\
      & 0.2 & 0.999644 & $2.2\times10^{-4}$ & 0.999718 & 0.999320 & 0.998943 & 0.98 & \\
      & 0.3 & 0.999201 & $5.0\times10^{-4}$ & 0.999368 & 0.998474 & 0.997617 & 0.73 & \\
      & 0.4 & 0.998583 & $8.9\times10^{-4}$ & 0.998881 & 0.997294 & 0.995756 & 0.41 & \\[3pt]
$\Bd$ & 0.1 & 0.999910 & $6.0\times10^{-5}$ & 0.999925 & 0.999818 & 0.999657 & 1.00 & 8.99\\
      & 0.2 & 0.999639 & $2.4\times10^{-4}$ & 0.999699 & 0.999278 & 0.998639 & 0.99 & \\
      & 0.3 & 0.999189 & $5.3\times10^{-4}$ & 0.999322 & 0.998388 & 0.996960 & 0.75 & \\
      & 0.4 & 0.998562 & $9.5\times10^{-4}$ & 0.998791 & 0.997155 & 0.994640 & 0.39 & \\[3pt]
$\Bs$ & 0.1 & 0.999910 & $5.9\times10^{-5}$ & 0.999926 & 0.999820 & 0.999660 & 1.00 & 8.95\\
      & 0.2 & 0.999641 & $2.4\times10^{-4}$ & 0.999703 & 0.999282 & 0.998652 & 0.98 & \\
      & 0.3 & 0.999193 & $5.3\times10^{-4}$ & 0.999332 & 0.998385 & 0.996995 & 0.75 & \\
      & 0.4 & 0.998569 & $9.4\times10^{-4}$ & 0.998815 & 0.997133 & 0.994712 & 0.38 & \\
\end{tabular}
\end{ruledtabular}
\end{table*}

\newchange{Fabrication variations cause the insertion loss to vary between TBUs. To assess how these variations affect fidelity, we performed Monte Carlo simulations in which each TBU's insertion loss was drawn independently from a Gaussian distribution $\mathcal{N}(\mu,\sigma^2)$ with a mean of $\mu=0.60$~dB. Following the convention in Table~\ref{tab:processor-model}, we describe the spread by $a=3\sigma$ and consider $a=0.1$, $0.2$, $0.3$, and $0.4$~dB. Each realization represents a possible device with a fixed loss pattern, and its fidelity $F_c$ is obtained by averaging Eq.~\eqref{eq:processor-fidelity} over the programmed time instants. We evaluated 100 realizations for each loss spread and each meson system, giving 1200 cases in total. The same underlying Gaussian draws were reused across all meson systems and rescaled for each value of $a$, allowing direct comparisons under matched loss patterns. With 100 realizations, the standard error of the mean fidelity $\langle F\rangle$ is approximately $6\%$ of the mean infidelity $1-\langle F\rangle$ at each loss spread.}

\begin{figure*}[p]
  \centering
  \includegraphics[width=\textwidth]{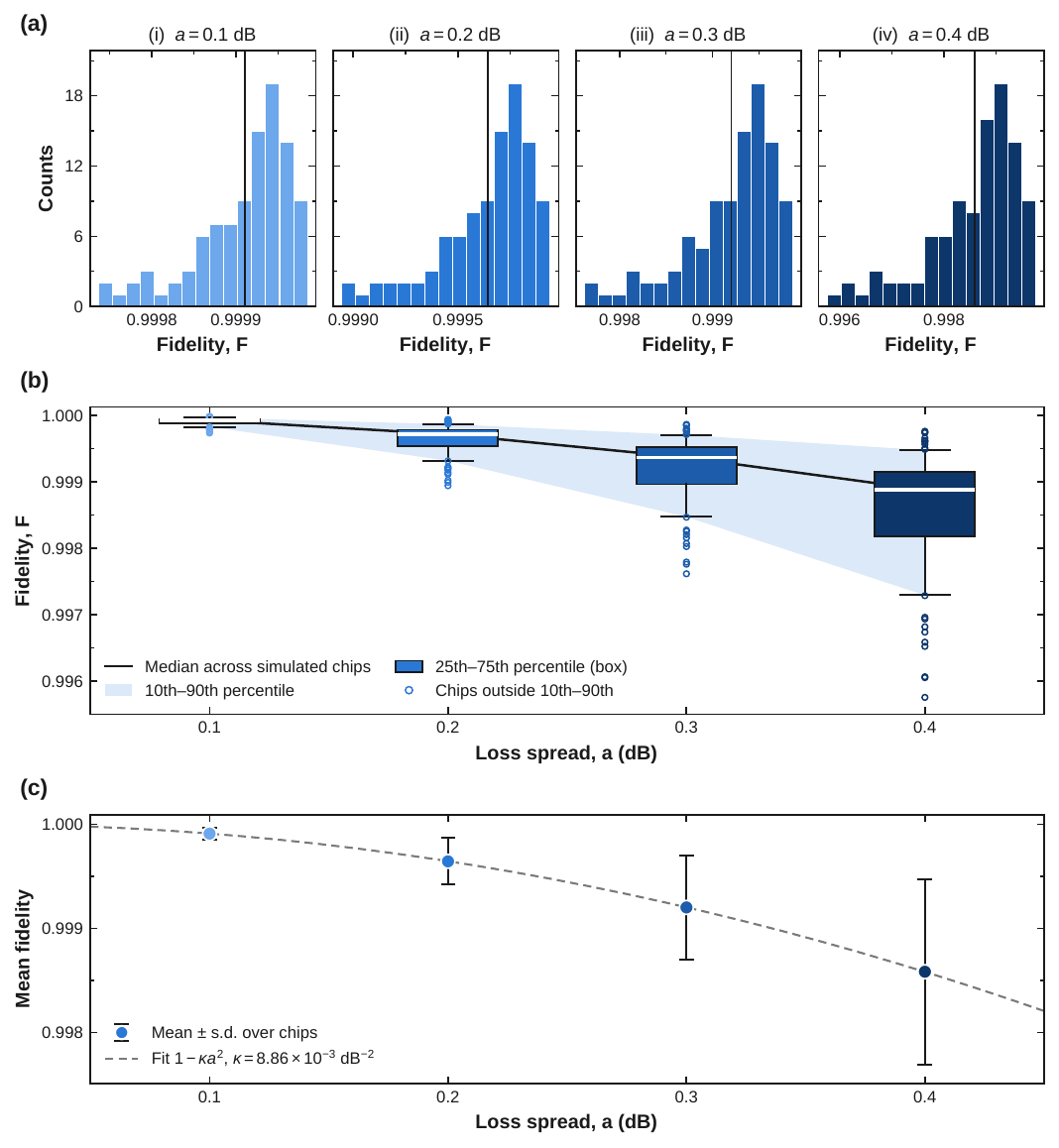}
  \caption{Monte Carlo analysis of the $\Kz$--$\Kzb$ simulation under random, nonuniform loss. Each of 100 simulated chips has independent Gaussian TBU insertion losses with mean $0.60$~dB and standard deviation $\sigma$, for loss spreads $a=3\sigma=0.1$, $0.2$, $0.3$, and $0.4$~dB (lighter to darker blue). The chip fidelity $F_c$ is the fidelity of Eq.~\eqref{eq:processor-fidelity} averaged over the programmed instants. (a)~Histograms of $F_c$ for (i)~$a=0.1$, (ii)~$0.2$, (iii)~$0.3$, and (iv)~$0.4$~dB; the vertical line marks the mean $\langle F\rangle$, and the fidelity scale differs between subpanels. (b)~Distribution of $F_c$ versus $a$: boxes span the 25th--75th percentiles with the median as a white bar, whiskers and the shaded band span the 10th--90th percentiles, open circles mark chips outside this range, and the solid line joins the medians. (c)~Mean fidelity $\langle F\rangle$ versus $a$ with error bars of $\pm1$ standard deviation over chips; the dashed curve is the least-squares fit $\langle F\rangle=1-\kappa a^2$ [Eq.~\eqref{eq:kappa}] with $\kappa=8.86\times10^{-3}$~dB$^{-2}$.}
  \label{fig:mc-K}
\end{figure*}

\begin{figure*}[p]
  \centering
  \includegraphics[width=\textwidth]{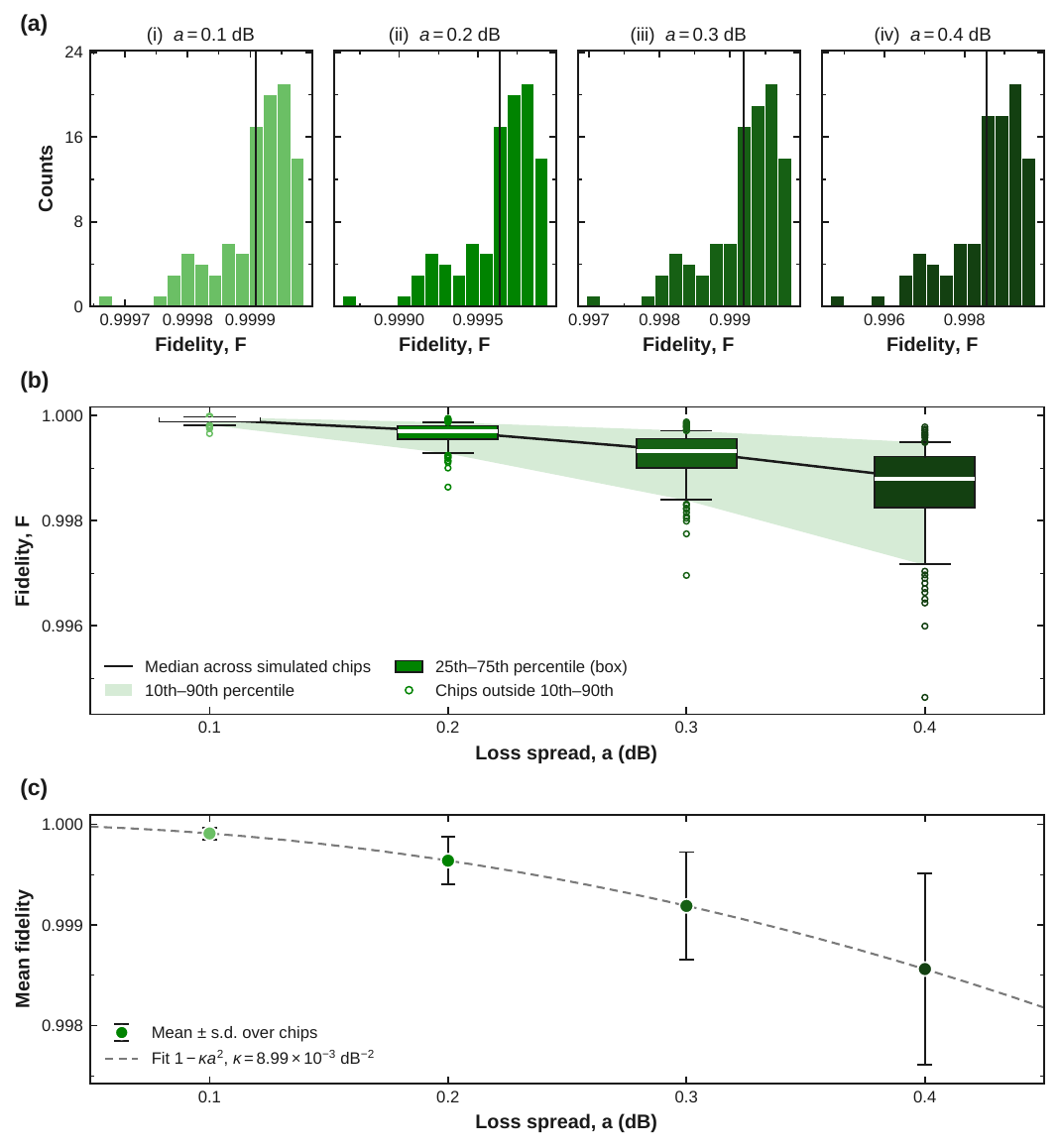}
  \caption{Monte Carlo analysis of the $\Bd$--$\Bdb$ simulation under random, nonuniform loss, on the same 100 simulated chips as Fig.~\ref{fig:mc-K}. Each chip has independent Gaussian TBU insertion losses with mean $0.60$~dB and standard deviation $\sigma$, for loss spreads $a=3\sigma=0.1$, $0.2$, $0.3$, and $0.4$~dB (lighter to darker green). The chip fidelity $F_c$ is the fidelity of Eq.~\eqref{eq:processor-fidelity} averaged over the programmed instants. (a)~Histograms of $F_c$ for (i)~$a=0.1$, (ii)~$0.2$, (iii)~$0.3$, and (iv)~$0.4$~dB; the vertical line marks the mean $\langle F\rangle$, and the fidelity scale differs between subpanels. (b)~Distribution of $F_c$ versus $a$: boxes span the 25th--75th percentiles with the median as a white bar, whiskers and the shaded band span the 10th--90th percentiles, open circles mark chips outside this range, and the solid line joins the medians. (c)~Mean fidelity $\langle F\rangle$ versus $a$ with error bars of $\pm1$ standard deviation over chips; the dashed curve is the least-squares fit $\langle F\rangle=1-\kappa a^2$ [Eq.~\eqref{eq:kappa}] with $\kappa=8.99\times10^{-3}$~dB$^{-2}$.}
  \label{fig:mc-Bd}
\end{figure*}

\begin{figure*}[p]
  \centering
  \includegraphics[width=\textwidth]{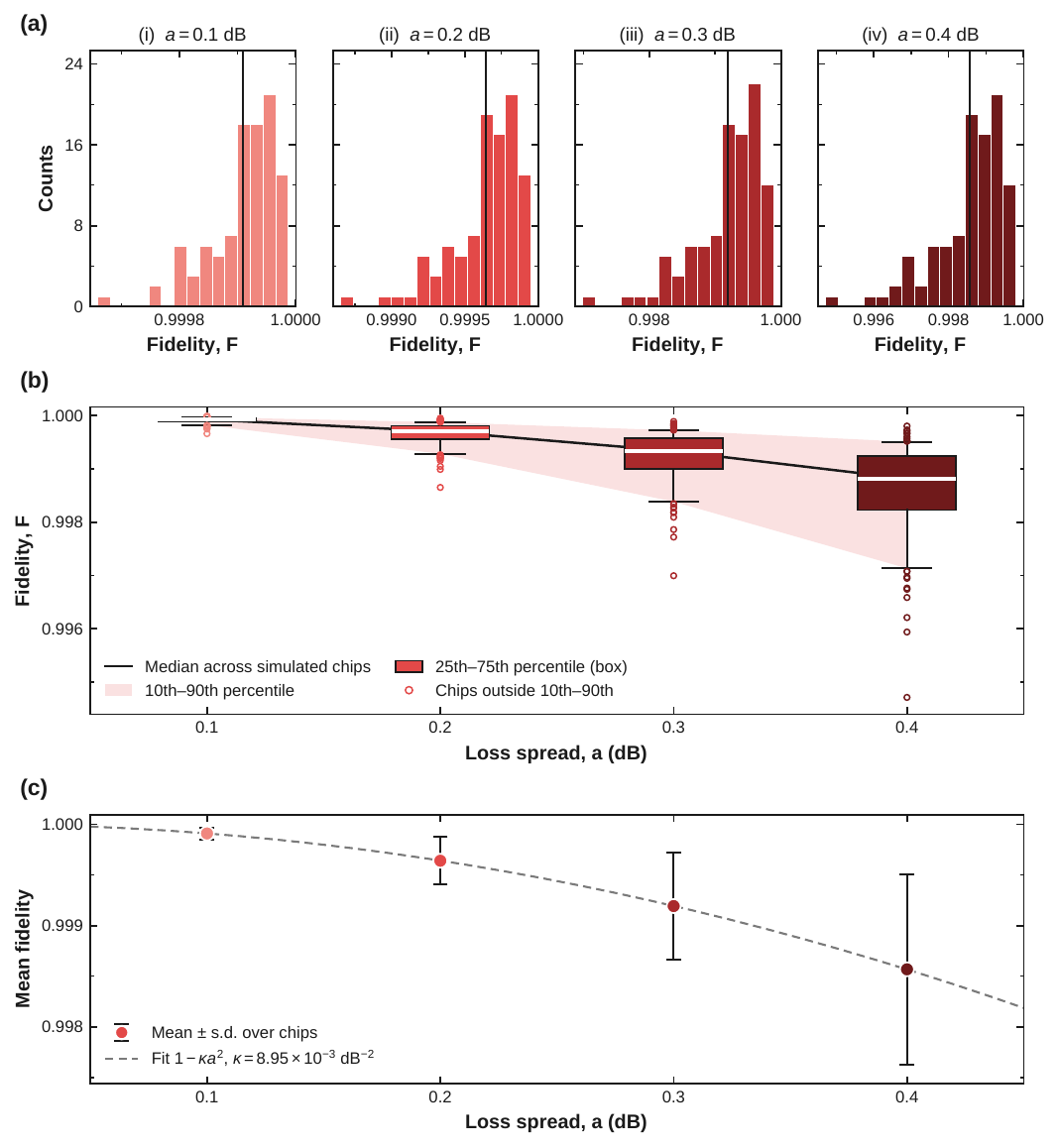}
  \caption{Monte Carlo analysis of the $\Bs$--$\Bsb$ simulation under random, nonuniform loss, on the same 100 simulated chips as Fig.~\ref{fig:mc-K}. Each chip has independent Gaussian TBU insertion losses with mean $0.60$~dB and standard deviation $\sigma$, for loss spreads $a=3\sigma=0.1$, $0.2$, $0.3$, and $0.4$~dB (lighter to darker red). The chip fidelity $F_c$ is the fidelity of Eq.~\eqref{eq:processor-fidelity} averaged over the programmed instants. (a)~Histograms of $F_c$ for (i)~$a=0.1$, (ii)~$0.2$, (iii)~$0.3$, and (iv)~$0.4$~dB; the vertical line marks the mean $\langle F\rangle$, and the fidelity scale differs between subpanels. (b)~Distribution of $F_c$ versus $a$: boxes span the 25th--75th percentiles with the median as a white bar, whiskers and the shaded band span the 10th--90th percentiles, open circles mark chips outside this range, and the solid line joins the medians. (c)~Mean fidelity $\langle F\rangle$ versus $a$ with error bars of $\pm1$ standard deviation over chips; the dashed curve is the least-squares fit $\langle F\rangle=1-\kappa a^2$ [Eq.~\eqref{eq:kappa}] with $\kappa=8.95\times10^{-3}$~dB$^{-2}$.}
  \label{fig:mc-Bs}
\end{figure*}

\newchange{Figures~\ref{fig:mc-K}--\ref{fig:mc-Bs} and Table~\ref{tab:mc} summarize the fidelity distributions across the simulated loss realizations. At $a=0.1$~dB, the fidelity exceeds $F_c=0.9996$ for every realization and all three meson systems, with a mean infidelity of approximately $9\times10^{-5}$. Increasing the spread to $a=0.4$~dB lowers the mean fidelity to about $0.9986$ and increases the standard deviation across realizations to approximately $9\times10^{-4}$. The lowest fidelity is about $0.995$. The distributions have a tail toward lower fidelity, which places the median above the mean at every loss spread. At $a=0.4$~dB, the median fidelity ranges from $0.9988$ to $0.9989$ across the three systems. Reusing the same underlying Gaussian draws allows the realizations to be compared directly as the loss spread increases. The histograms retain similar shapes, consistent with an approximately quadratic scaling of the infidelity.

As shown in panels (c) of Figs.~\ref{fig:mc-K}--\ref{fig:mc-Bs}, the mean infidelity follows
\begin{equation}
  1-\langle F\rangle \simeq \kappa a^2
  = 9\kappa\sigma^2,
  \label{eq:kappa}
\end{equation}
where the fitted values of $\kappa$ are $(8.86,\,8.99,\,8.95)\times10^{-3}$~dB$^{-2}$ for $\Kz$, $\Bd$, and $\Bs$, respectively. All three systems therefore give approximately $1-\langle F\rangle=0.080\,\sigma^2$ when $\sigma$ is expressed in decibels. Allowing the exponent to vary in a power-law fit gives $1.996$ for each system, and the relative residuals of the quadratic fit remain below $0.6\%$. Similar scaling is observed for individual loss realizations. Relative to $a=0.1$~dB, the infidelity increases by approximately $3.99$, $8.97$, and $15.9$ at $a=0.2$, $0.3$, and $0.4$~dB. These factors are close to the values of $4$, $9$, and $16$ expected from an exact quadratic dependence.

A perturbative analysis explains both the quadratic dependence and the magnitude of $\kappa$. We express the insertion loss of TBU $k$ as $L_k=\mu+\delta_k$, where $\delta_k$ is its deviation from the mean. Every optical path crosses nine TBUs, so the common loss $\mu$ contributes the same attenuation to all paths and cancels from the normalized fidelity in Eq.~\eqref{eq:processor-fidelity}. To first order in the loss variations,
\[
  S \simeq 10^{-9\mu/20}
  \left(\widehat{\mathcal{U}}-c\sum_k\delta_k A_k\right),
  \qquad c=\frac{\ln 10}{20}.
\]
Here, $A_k=B_kP_kC_k$ describes the contribution from the modes passing through TBU $k$. The matrices $C_k$ and $B_k$ describe the ideal unitary evolution before and after the attenuation at that unit, with $\widehat{\mathcal{U}}=B_kC_k$. The projector $P_k$ selects the $r_k\in\{1,2\}$ modes that pass through it.

A perturbation proportional to $\widehat{\mathcal{U}}$ changes only the overall amplitude and phase, leaving the normalized fidelity unchanged. The leading reduction in fidelity therefore comes from the orthogonal component
\[
  A_k^\perp
  =A_k-\frac{1}{N}
  \operatorname{tr}\!\left(\widehat{\mathcal{U}}^\dagger A_k\right)
  \widehat{\mathcal{U}},
\]
which gives
\[
  1-F \simeq
  \frac{c^2}{N}
  \left\|\sum_k\delta_k A_k^\perp\right\|_F^2.
\]
Here, $N=4$ and $\|\cdot\|_F$ denotes the Frobenius norm. For independent, zero-mean deviations with variance $\sigma^2$, the cross terms between different TBUs vanish upon averaging. Using
$\|A_k\|_F^2=\operatorname{tr}(\widehat{\mathcal{U}}^\dagger A_k)=r_k$
then gives
\begin{equation}
  1-\langle F\rangle \simeq
  \frac{c^2\sigma^2}{N}
  \sum_k r_k\left(1-\frac{r_k}{N}\right).
  \label{eq:meanF}
\end{equation}

At leading order, the mean infidelity therefore depends on how many modes pass through each TBU and is independent of the programmed unitary. In this mesh, 30 TBUs carry light. The six mixing elements in Eq.~\eqref{eq:processor-layout} each carry two modes, while the remaining 24 units each carry one. Thus, $\sum_k r_k=36$, consistent with four modes each passing through nine TBUs, and $\sum_k r_k(1-r_k/4)=24$. Equation~\eqref{eq:meanF} consequently predicts
\[
  1-\langle F\rangle \simeq
  6c^2\sigma^2 \approx 0.0795\,\sigma^2,
\]
corresponding to $\kappa=6c^2/9\approx8.84\times10^{-3}$~dB$^{-2}$. This prediction differs from the fitted coefficients by only $0.3\%$ to $1.7\%$, within their statistical uncertainty. As a further check, we evaluated Eq.~\eqref{eq:processor-fidelity} for random target unitaries using an equivalent network with six two-mode units and 24 one-mode units, arranged so that every path contains the same number of units. The resulting mean infidelity agrees with Eq.~\eqref{eq:meanF} to within approximately $1\%$.

Equation~\eqref{eq:kappa} provides an estimate of the allowable variation in TBU insertion loss under the assumed model. A mean fidelity of at least $0.999$ requires approximately $a\lesssim0.33$~dB, or $\sigma\lesssim0.11$~dB. Raising this target to $0.9999$ reduces the allowable spread to $a\lesssim0.105$~dB, or $\sigma\lesssim0.035$~dB. The fraction of individual realizations meeting the threshold $F_c\geq0.999$ also decreases as the spread increases. Across the three meson systems, this fraction is $100\%$ at $a=0.1$~dB, $98\%$ to $99\%$ at $0.2$~dB, $73\%$ to $75\%$ at $0.3$~dB, and $38\%$ to $41\%$ at $0.4$~dB, as reported in Table~\ref{tab:mc}.}

\section{Conclusion}
\label{sec:conclusion}

We have presented a protocol that maps the non-unitary evolution of a neutral-meson pair onto a reprogrammable linear-optical processor and have analyzed its performance with circuit-level simulations of a hexagonal waveguide mesh. The effective Hamiltonian is reconstructed exactly from measured observables, its propagator is dilated with \newchange{a single ancilla that encodes the total decay probability}, and the resulting four-mode unitary is compiled into a Clements interferometer at every time step. Reprogrammed only through its control voltages, one fixed device reproduces the strongly lifetime-asymmetric kaon system, the slowly oscillating $B_d$ system, and the rapidly oscillating $B_s$ system with the same mean fidelity of about $0.9998$, although the three cases span a factor of 35 in $x$ and two orders of magnitude in $\Gamma$. This uniformity reflects the structure of the problem: once time is measured in units of $1/\Gamma$, the target transformation depends only on $x$, $y$, and $|q/p|$. Loss enters in two distinct ways. Because every optical path crosses the same number of tunable units, the common part of the insertion loss reduces only the throughput and leaves both the simulated probabilities and the fidelity unchanged, whereas the unit-to-unit variation lowers the fidelity at second order, with a mean effect set by the routing of light through the mesh rather than by the simulated dynamics. This explains why the loss tolerance is the same for the three mesons and why a leading-order estimate reproduces the fitted coefficient to within $2\%$. \newchange{For the loss spreads assumed in the circuit model, the loss-induced infidelity is of order $10^{-4}$ under ideal phase settings and coupler splitting ratios.} \newchange{The dedicated ancillary outputs encode the total decay probability without postselection on the ancilla outcome. Individual outputs are not identified with physical decay channels, and the reported probabilities remain conditional on optical collection.} Programmable photonic meshes therefore offer a practical route to the quantum simulation of unstable-particle dynamics. Natural next steps are single-photon experiments on fabricated meshes, the inclusion of phase and coupler errors in the same Monte Carlo framework, and higher-extinction architectures for the weakly mixing $D^0$ system.

\cpsuggestion{10}{\textbf{Abstract and conclusion.} With the uploaded generator, describe kaon and $B_s$ mixing as CP violating and $B_d$ mixing as CP conserving. If $B_d$ is rerun, update any affected numerical summaries. The alternative $B_d$ and $B_s$ central values define CP-violating benchmarks, but their measured $|q/p|$ values are compatible with unity; do not present them as significant experimental evidence for CP violation in mixing.}

\begin{acknowledgments}
The authors used Claude (Opus 5.5; Anthropic) solely for language editing to improve the grammar, clarity, and tone of the manuscript. All AI-assisted suggestions were reviewed by the authors, who take full responsibility for the final content of the manuscript.
\end{acknowledgments}

\bibliography{references}

\end{document}